\documentclass{ws-mpla}

\begin{document}

\markboth{Saneesh Sebastian and V C Kuriakose}
{Scattering of Scalar field around an Extended black hole in F(R) gravity}

\catchline{}{}{}{}{}

\title{Scattering of Scalar Field by an Extended Black Hole in F(R) gravity}

\author{\footnotesize  Saneesh Sebastian\footnote{
Typeset names in 8 pt Times Roman, uppercase. Use the footnote to
indicate the present or permanent address of the author.}}

\address{Department of Physics, Cochin University of Science and
Technology, Kochi 682022, India
saneeshphys@cusat.ac.in}

\author{V C Kuriakose}

\address{Department of Physics, Cochin University of Science and
Technology, Kochi 682022, India\\
vck@cusat.ac.in
}

\maketitle

\pub{Received (Day Month Year)}{Revised (Day Month Year)}

\begin{abstract}
In this work we have studied the scattering of scalar field around
an extended black hole in F(R) gravity using WKB method. We have
obtained the wave function in different regions such as near the
horizon region, away from horizon and far away from horizon and  the
absorption cross section are calculated. We
find that the absorption cross section is inversely proportional to
the  cube  of Hawking temperature.  We  have  also  evaluated  the
Hawking  temperature  of  the  black  hole  via  tunneling  method.

\end{abstract}

\ccode{PACS Nos.: 04.70.Dy, 04.70.-s }

\section{Introduction}

The late time acceleration of the universe\cite{pm,rs} introduced
new problems in gravitational physics. Modification of Einstein's
general theory of relativity is required to explain the  late time
cosmic acceleration. There are two methods for explaining these new
observations, one method is to introduce  the  concept of dark
energy to provide  the  necessary  force  for  acceleration and the
other method is to modify Einstein's equation of General Theory of
Relativity(GTR). F(R) model gravity is an attempt to modify GTR
\cite{cf}, for a review see reference [4]\cite{tv}. There exist black hole solutions in this new formulation
of gravity also\cite{jdb,ti}.
    \paragraph{} There are a number of modified gravity theories and F(R) theory is the simplest one. In
Einstein's formulation the Ricci scalar is used as the Lagrangian
density. In F(R) theory,  we use a general function of Ricci scalar
as the Lagrangian density. Due to the appearance of the function of
the Ricci scalar the resulting field equation is fourth order, while
the field equation in Einstein's general relativity is second order.
This makes the study of  F(R) field equation more difficult but
attempts have been made to obtain cosmological solution and Black
hole solution in this new frame  work.

 \paragraph{}Black holes are objects capable of absorbing all particles and radiation that
 enter the event horizon. But in 1970s Bekenstein\cite{bek} and
 Hawking\cite{swh} developed the black hole thermodynamics which indicates that Black
Holes have a characteristic temperature and entropy. Semi-classical
theory shows that Black Holes can emit particles\cite{sw}. This is
called particle creation by Black Holes. These studies led  to  a
new area  of  study  called  thermodynamics  of  black  holes  and
this area has attracted  attention  of  many. Classically all
matter entering  the  event  horizon get  absorbed  by  the black
hole. But  quantum  mechanically  there  is  a    probability of
matter and/or  fields   getting scattered. This shows a non-zero
reflectivity at the horizon. Using this assumption a number of works
have appeared in the literature studying Black Hole
scattering\cite{fm}.
\paragraph{} The paper is organized as follows. In section 2 we give the basic theory of black hole scattering
assuming  that  black  hole  is surrounded    by  a  scalar  field
which  obeys  the Klein-Gordon equation.   In section 3 we study the
the wave function of scattering field in the vicinity of the
horizon.  The solution of wave equation in a region $r$ greater than
$r_{h}$ is studied in the section 4. In section 5 the solution at
far away from horizon is studied. The absorption cross section is
studied in section 6.  In section 7 we  calculate  the  Hawking
temperature via  tunneling  method.  Conclusion of the present study
is given in section 8.
\section{Theory of black hole scattering}
The action for  the  gravity  in  F(R)  theory can  be written as
\begin{equation}
 S=\frac{1}{2\kappa^{2}}\int d^{4}x \sqrt{-g} F(R),
\end{equation}
where $g$ is the determinant of the metric and $F$ is a general
function of the Ricci scalar $R$. We start with a static spherically
symmetric solution of the form \cite{seb}
\begin{equation}
 ds^{2}=-e^{2\beta (r)}B(r)dt^{2}+\frac{dr^{2}}{B(r)}+r^{2}d\Omega^{2}.
\end{equation}
Following  reference.[11], we  assume  that $\beta$ is constant and
$F(R)=\sqrt{R+6C_{2}}.$
 We get a solution  of  the  form
\begin{equation}
 -B \left(r\right) d t^2 + \frac{dr^2}{B \left(r\right)} + r^2 d \Omega^2,
\end{equation}
with $B \left(r\right)$ equal to,
\begin{equation}
 B \left(r\right)=1-\frac{c_1}{r^2}+c_2 r^2.
\end{equation}
The $c_{2}$ term in the solution represents a cosmological constant
term.   Assuming the cosmological constant term is zero ie. $c_2=0$
and $c_1=2 \alpha m$,  we  get,

\begin{equation}
 ds^{2}=-(1-\frac{2m\alpha}{r^{2}})dt^{2}+\frac{dr^{2}}{(1-\frac{2m\alpha}{r^{2}})}+r^{2}{d\theta^{2}}+r^{2} sin^{2}\theta d\phi^{2},
\end{equation}
where all the symbols except $\alpha$ have the usual meaning.
$\alpha$ is a constant parameter having  the dimension of length.
The metric has  singularities  at both $r=\sqrt{2\alpha m}$ and
$r=0$. Since $\alpha$ is a constant parameter forms a black hole solution
with horizon at $r=\sqrt{2\alpha m}$. We  will  now  invert the
metric into tortoise co-ordinate as
\begin{equation}
 ct=\pm r \pm \sqrt{2\alpha m}~  arctanh[\frac{r}{\sqrt{2\alpha m}}]+
 constant,
\end{equation}
where the minus sign is for ingoing photon and positive sign is for
outgoing photon. Proceeding as in GTR, we first obtain a new
coordinate $p$ with $p$ as
\begin{equation}
 ct=-r-\sqrt{2\alpha m}~arctanh[\frac{r}{\sqrt{2\alpha m}}]+p.
\end{equation}
With the null coordinate $p$, the metric simplyfy Eq.(5) can be written as
\begin{equation}
 ds^{2}=\left(1-\frac{2\alpha
 m}{r^{2}}\right)dp^{2}-2dpdr-r^{2}d\Omega^{2}.
\end{equation}
Now defining a time like coordinate $t'$ as $ ct'=p-r $ so that we
get the Eddington-Finkelstein coordinate metric as
\begin{equation}
 ds^{2}=c^{2}\left(1-\frac{2\alpha m}{r^{2}}\right)dt'^{2}-\frac{4m\alpha c}{r^{2}}dt'dr-\left(1+\frac{2\alpha m}{r^{2}}\right)dr^{2}-r^{2}d\Omega^{2}
\end{equation}
The Eddington-Finkelstein coordinate metric for   the extended black
hole has  the same form as Schwarzschild metric and  thus this
metric can have a regular null surface at $r=\sqrt{2\alpha m}$.

    We  now place  the Black hole  immersed in a massive scalar field, the field equation in this back ground can
be described by the Klein-Gordon equation\cite{sv}

\begin{equation}
(\square+\mu^{2})\Phi=0.
\end{equation}

In curved space-time this equation can be written in the following form,
\begin{equation}
\frac{1}{\sqrt{-g}}\partial_{
\mu}(\sqrt{-g}{g}^{\mu\nu}\partial_{\nu}\Phi)+\mu^{2}\Phi)=0.
\end{equation}
Writing this equation in the spherical polar coordinate we find\cite{wu}
\begin{equation}
\frac{1}{r^{2}(sin\theta)}[(\partial_{t}(r^{2}(sin\theta)) \frac{1}{(1-\frac{2m\alpha}{r^{2}})}\partial_{t})-(\partial_{r}{r^{2}sin\theta}{(1-\frac{2m\alpha}{r^{2}})}\partial_{r})-(\partial_{\theta}{r^{2}sin\theta}\frac{1}{r^{2}}\partial_{\theta})-(\frac{1}{sin\theta}{\partial_{\phi}}^{2})]\Phi=0.
\end{equation}

In order to separate this equation into radial and the angular parts, we assume that $\Phi$ is of the form

\begin{equation}
\Phi(r,t)=exp(-i\epsilon t)\phi_{l}(r) Y_{lm}(\theta, \phi),
\end{equation}
where $\epsilon$ is the energy, $l$ and $m$ are the angular momentum
and its projection respectively. We consider only the radial part
and it takes the following form
\begin{equation}
[\frac{1}{r^{2}}\partial_{r}(r^{2}-2\alpha m)\partial_{r}+\frac{r^{2}\epsilon^{2}}{(r+\sqrt{2\alpha m})(r-\sqrt{2\alpha m})}-\frac{l(l+1)}{r^{2}}+\mu^{2}]\phi_{l}(r)=0.
\end{equation}

\section{Solution of wave equation in the vicinity of horizon}
We use WKB approximation for solving the radial scattering equation
and assume\cite{yk} a  solution  of  the  form,
\begin{equation}
\phi_{l}(r)=exp(-i\int k(r)dr).
\end{equation}
Using  Eq. 15  in  Eq.7 we  get,
\begin{equation}
[\frac{\epsilon^{2}}{1-\frac{2\alpha m}{r^{2}}}-\frac{l(l+1)}{r^{2}}+\mu^{2}]+[-k^{2}(1-\frac{2\alpha m}{r^{2}})]=0.
\end{equation}
On rearranging we find,
\begin{equation}
k^{2}=(1-\frac{2\alpha m}{r^{2}})^{-2}[\epsilon^{2}-(\frac{l(l+1)}{r^{2}}+\mu^{2})(1-\frac{2\alpha m}{r^{2}})].
\end{equation}
Since we   are  considering  a  situation  where $r$ approaches
$r_{h}$ and  hence  we  need  only  consider  the  $\epsilon^{2}$
 term  in  the  square  bracket.  Then $k$ is given by,
\begin{equation}
k=(1-\frac{2\alpha m}{r^{2}})^{-1}\epsilon.
\end{equation}
and
\begin{equation}
 (1-\frac{2m\alpha}{r^{2}})=\frac{1}{r^{2}}(r^{2}-2\alpha m)=\frac{1}{r^{2}}{(r+\sqrt{2\alpha m})(r-\sqrt{2\alpha m})}
\end{equation}
When $r \rightarrow r_{h}$, $k$  can  be  written   in  a  compact
form  as,
\begin{equation}
k=\frac{\xi}{(r-r_{h})}.
\end{equation}
where $\xi$ is
\begin{equation}
\xi=\frac{r^{2}_{h}\epsilon}{r+r_{h}}.
\end{equation}
Hence from our above assumption, the wave function as $r\rightarrow r_{h}$ becomes
\begin{equation}
\phi_{l}=e^{i\int{\frac{\xi}{r-r_{h}}}}.
\end{equation}
which can be integrated to the following form,
\begin{equation}
\phi_{l}=e^{i\xi ln(r-r_{h})}.
\end{equation}
Thus  the  wave  function   in  the vicinity  of the  event horizon
may  be  written  as,
\begin{equation}
\phi_{l}=e^{\pm i\xi ln(r-r_{h})}.
\end{equation}
Now  we  will  consider   the  scalar wave  approaching  the  black
hole  horizon  and  using  Eq. (24)  the  wave  function  in  the
vicinity   of  the black  hole  horizon  can  be  written,  assuming
the field  gets  reflected  at  the  black  hole  horizon,  as
\begin{equation}
\phi_{l}=e^{-{i ln(r-r_{h})}} +  |R|e^{+{i ln(r-r_{h})}}
\end{equation}
where  $R$   represents   the  reflection  coefficient. If  $R \neq
0$,  there  is  a definite probability   for  the  incident  wave  to
get   reflected  at  the  horizon.
\section{Solution of wave equation in a region r greater than $r_{h}$.}
In  this  section  we  consider  a  situation  where  the  field  is
sufficiently  away  from the event  horizon. We  also  assume  a
situation  that  the energy and  momentum, $\epsilon$ and $\mu$ of
the  field  are  very small and  they  can  be  neglected. For
s-wave Eq. (14)   now  takes the  form,
\begin{equation}
\frac{1}{r^{2}}\partial_{r}(r^{2}-2\alpha m \partial_{r}\phi_{0})=0.
\end{equation}
From which we obtain, after some calculations, the following
equation

\begin{equation}
ln\phi'_{0}(r)=-ln(2r(r+\sqrt{2\alpha m})(r-\sqrt{2\alpha m})),
\end{equation}
and the wave function is obtained as,
\begin{equation}
\phi_{0}=Cln(\frac{r-r_{h}}{r}).
\end{equation}
Comparing the solutions for regions 1 and 2,
we have in Region 1 the wave function as
\begin{equation}
\phi_{l}=e^{i\xi ln(r-r_{h})},
\end{equation}
which contains both incoming and outgoing waves. For  $s$-wave, we can
rewrite  the  solution  for  Region 1,  as
\begin{equation}
\phi_{0}={1-i\xi ln (r-r_{h})+R(1+i\xi ln(r-r_{h}))}.
\end{equation}
Where R is the reflection coefficient\cite{ku1}. 
This can be further simplified as
\begin{equation}
\phi_{0}=-i\xi ln (r-r_{h})(1-R)+(1+R).
\end{equation}
Thus in Region 2  also,  we  can  write   the  $s$ -  wave solution
as
\begin{equation}
\phi_{0}=\alpha ln\frac{r-r_{h}}{r}+\beta,
\end{equation}
where $\alpha=i\xi(1-R)$ and $\beta=(1+R)$.

  In  the next  section  we  will  study   the   behaviour  scalar
  field in  a  region sufficiently  far  away  from  the  horizon.

\section{Solution of wave equation far away from the horizon}

In this region, the radial part  of  the  wave equation can be
written as
\begin{equation}
\phi_{l}\textquotedblright+\frac{2}{r} \phi_{l}\textquoteright+p^{2}\phi_{l}=0,
\end{equation}
with p is the linear momentum associated with scalar field and is $p^{2}=\epsilon^{2}-\mu^{2}$.

Using Frobenius method we can solve the above equation. The solution is given by

\begin{equation}
\phi_{l}=\frac{1}{r}(A_{l}e^{iz}+B_{l}e^{-iz}),
\end{equation}
with $z=pr.$
This can be again simplified as
\begin{equation}
\phi_{l}=\frac{1}{r}(a F(r)+b G(r)),
\end{equation}
where $F(r)=sin(pr)$ and $G(r)=cos(pr)$. At the boundary between
Region 2 and Region 3, where $pr\ll1$ then we can expand both sine
and cosine terms and need to take only first terms; then
$F(r)\approx pr$ and $G(r)\approx 1$ and  the   $s$ - wave  solution
is  given  by,
\begin{equation}
\phi_{0}=ap+\frac{b}{r},
\end{equation}

and in Region 2, $\phi_{0}$ is given by
\begin{equation}
\phi_{0}= -\alpha \frac{r_{h}}{r} +\beta
\end{equation}
From region 1 we get the wave function finally as
\begin{equation}
\phi_{0}=i\xi (1-R)\frac{r_{h}}{r} +(1+R).
\end{equation}
Comparing equations (36) and (37) we can obtain $a$ and $b$ as
$a=\frac{1+R}{p}, b=i\xi r_{h}(1-R)$. Now  we  will  calculate   the
absorption  coefficient,  assuming  the  wave  gets  reflected   at
the  horizon.

\section{Absorption cross section}
The absorption cross section can be evaluated using the above data.
We  now  consider the   solution  of  Eq.(32) given  by  Eq.(33).
The scattering matrix element is  defined  as
\begin{equation}
S_{l}=(-1)^{l+1}\frac{A_{l}}{B_{l}} e^{2i\delta_{l}},
\end{equation}
where  $\delta_{l}$ is the phase shift corresponding to angular
momentum $l$.

We take the low energy limit so that $l=0$. Using  Eqs. (33), (35)
and (37),  we   can  obtain,  for  $s$ - waves,
$A_{0}=\frac{a+ib}{2i}$ and $B_{0}=\frac{-a+ib}{2i}$. Substituting
the values of $a$ and $b$ we get $S_{0}$ as
\begin{equation}
S_{0}=\frac{(1+R)-\xi pr_{h}(1-R)}{(1+R)+\xi pr_{h}(1-R)}.
\end{equation}
Defining $\eta=\frac{1-R}{1+R}$
the equation of $S_{0}$ takes the form
\begin{equation}
S_{0}=\frac{1-\xi pr_{h}\eta }{1+\xi pr_{h}\eta}.
\end{equation}
The absorption cross section $\sigma_{abs}$, is then given by
\begin{equation}
\sigma_{abs}=\frac{\pi }{p^{2}}(1-S_{0}^{2}).
\end{equation}
Using the above equations and also using the relation $p=\epsilon v$ we obtain finally
\begin{equation}
\sigma_{abs}=\frac{2\pi^{2}\epsilon r_{h}^{3}}{v}.
\end{equation}
Since $r_{h}$ is inversely proportional to the Hawking temperature $T_{H}$, we can see that absorption cross section is 
inversely related to Hawking temperature. 
\section{Hawking temperature via tunneling}
Now  we  will   determine the Hawking temperature using  tunneling
mechanism\cite{pw}. This mechanism has been used by many authors\cite{km,sh,rb} 
for determining the Hawking radiation of black holes in Einstein gravity.
Even though the complete properties of Hawking 
radiation can be obtained using quantum field theory in curved space-time, 
 the tunneling mechanism gives a simple understandable picture.  
 According to this picture the radiation arises
by a process similar to electron-positron pair creation in a
constant electric field. Using tunneling picture of black hole radiation we can have a direct
semi-classical derivation of black hole radiation. There are two different schemes
 in tunneling approach, first is the radial null geodesic method and the other is 
 Hamilton-Jacobi method. Here we use the radial null geodesic method. The metric in
Eq.(5) can be transformed in to Painleve\cite{pp} like coordinate system to remove the singularity 
in the original extended metric as,
\begin{equation}
ds^{2}=-\left(1-\frac{2\alpha m}{r^{2}}\right)dt^{2}+2\sqrt{\frac{2\alpha m}{r^{2}}}dtdr+dr^{2}+r^{2}d\Omega^{2}.
\end{equation}
The radial null geodesics are
\begin{equation}
 \frac{dr}{dt}={\dot{r}}=\pm1-\sqrt{\frac{2\alpha m}{r^{2}}}.
\end{equation}
The typical wavelength of the radiation is of the order of the 
size of the black hole and hence when the outgoing wave is traced back towards the horizon 
its wavelength as measured by local  observers, is  
blue shifted. Near the horizon, radial wave number approachs to infinity and we can use  WKB
approximation to study for the particle tunneling . 
We start with original action 
\begin{equation}
 S=\int p(r)dr.
\end{equation}
Using Hamilton's equation of motion $p(r)=\frac{dH}{\dot{r}}$ we can write the imaginary part of the action as  
\begin{equation}
 Im S= Im\int_{r_{in}}^{r_{out}}p_{r}dr=Im \int_{m}^{m-\omega}\int_{r_{in}}^{r^{out}}\frac{dr dH}{\dot{r}}
\end{equation}

 Using Eq.(45), the imaginary part of the action S can be written as  

\begin{equation}
Im S=Im\int_{0}^{\omega}\int_{r_{in}}^{r_{out}}\frac{dr d\left(-\omega\right)}{\left(1-\sqrt{\frac{2\alpha \left(m-\omega\right)}{r^{2}}}\right)}.
\end{equation}
where $\omega$ is the frequency of out going particle. Eq.(48) can be integrated to obtain
\begin{equation}
ImS=\frac{4\pi\sqrt{2\alpha}}{3}\left(m-\omega\right)^{\frac{3}{2}}.
\end{equation}
Since $m$ is much greater than $\omega$ we can apply the binomial
expansion to get
\begin{equation}
Im S= \frac{4\pi\sqrt{2\alpha m}\ m}{3}\left(1-\frac{3\omega}{2m}\right).
\end{equation}
now the Semi-classical emission rate can be written as
\begin{equation}
\Gamma\sim e^{-2ImS}\sim exp\left(\frac{4\pi\sqrt{2\alpha m}\ m}{3}\left(\frac{3\omega}{2m}\right)\right)
\end{equation}
from which we get
\begin{equation}
 T_{H}=\frac{1}{2\pi \sqrt{2\alpha m}}.
\end{equation}
There exist higher order corrections of $\omega$ due to the conservation of energy, but for the first order calculation it is neglected.
It is possible to find out the frequency dependent transmission coefficient or graybody factors in this tunneling scenario.
\section{Conclusion}
In this paper we have studied the scattering properties of extended
black holes in F(R) theory. We have obtained the scattered wave
equation in both regions near the horizon and away from horizon.
Using the scattering method we have obtained the absorption cross section
and  also  calculated the  Hawking  temperature  of  the  black
hole via tunneling method.  
\section*{Acknowledgments}
The authors are thankful to the reviewer for useful comments.
SS wishes to thank CSIR, New Delhi for financial support under CSIR-SRF scheme. VCK is
thankful to UGC, New Delhi for financial support through a
Major Research Project and wishes to acknowledge Associateship of IUCAA, Pune, India



\begin{thebibliography}{0}
\bibitem{pm}  S. Perlmutter et. al. Astrophys. J. \textbf{517} 565 (1999)
\bibitem{rs}  A. G. Riess et al., Astronom. J. \textbf{116} 1009 (1998)
\bibitem{cf} S. Capozziello and V. Faraoni Beyond Einstein Gravity  Springer (2011)
\bibitem{tv} T. P. Sotiriou and V. Faraoni Rev. Mod. Phys.\textbf{82} 451 (2010)
\bibitem{jdb} T. Clifton and J.D. Barrow Phys. Rev. D \textbf{72} 103005 (2005)
\bibitem{ti}  T. Multamaki and I. Vilja, Phys. Rev. D \textbf{74} 064022 (2006)
\bibitem{bek} J. D. Bekenstein, Lett . Nuovo Cimento Soc. Ital. fis. \textbf{11} 467 (1974)
\bibitem{swh} S. W. Hawking and D. N. Page Commun. Math. Phys. \textbf{87} 577-588 (1983)
\bibitem{sw}  S. W. Hawking Commun. Math. Phys. \textbf{43} 199-220 (1975)
\bibitem{fm}  J. A. H. Futterman, F. A. Handler, and R. A.Matzner, Scattering from Black Holes (Cambridge University Press, New York, 1988)
\bibitem{seb} L. Sebastiani and S. Zerbini. Eur.Phys.J. C \textbf{71} 15915 (2011)
\bibitem{wu}  W. G. Unruh, Phys Rev D \textbf{14}, 3251(1976)
\bibitem{sv}  Sini. R and V. C. Kuriakose Int. J. Mod. Phys.D \textbf{16} 105-116(2007)
\bibitem{yk}  M. Y. Kuchiev and V. V. Flambaum Phys. Rev. D \textbf{70} 044022 (2004)
\bibitem{ku1} M. Y. Kuchiev Phys. Rev. D \textbf{69}, 124031 (2004)
\bibitem{pw}  M. K. Parikh  and F. Wilczek Phys. Rev. Lett.\textbf{85} 24 (2000)
\bibitem{km}  K. Matsuno Phys. Rev. D \textbf{83} 064016 (2011)
\bibitem{sh}  S. H. Mehdipour Phys. Rev. D \textbf{81} 124049 (2010)
\bibitem{rb}  R. Banerjee et.al. Phys. Rev. D \textbf{77} 124035 (2008)
\bibitem{pp} P. Painleve C. R. Acad. Sci. (Paris) \textbf{173}, 677 (1921)
\end{thebibliography}
\end{document}